# A New Probabilistic V-Net Model with Hierarchical Spatial Feature Transform for Efficient Abdominal Multi-Organ Segmentation

Minfeng Xu, Heng Guo, Jianfeng Zhang, Ke Yan, and Le Lu✉

DAMO Academy, Alibaba Group, Hangzhou, 311121, China
{tiger.lelu}@gmail.com

**Abstract.** Accurate and robust abdominal multi-organ segmentation from CT imaging of different modalities is a challenging task due to complex inter- and intra-organ shape and appearance variations among abdominal organs. In this paper, we propose a probabilistic multi-organ segmentation network with hierarchical spatial-wise feature modulation to capture flexible organ semantic variants and inject the learnt variants into different scales of feature maps for guiding segmentation. More specifically, we design an input decomposition module via a conditional variational auto-encoder to learn organ-specific distributions on the low dimensional latent space and model richer organ semantic variations that is conditioned on input images. Then by integrating these learned variations into the V-Net decoder hierarchically via spatial feature transformation, which has the ability to convert the variations into conditional Affine transformation parameters for spatial-wise feature maps modulating and guiding the fine-scale segmentation. The proposed method is trained on the publicly available AbdomenCT-1K dataset and evaluated on two other open datasets, i.e., 100 challenging/pathological testing patient cases from AbdomenCT-1K fully-supervised abdominal organ segmentation benchmark and 90 cases from TCIA$^+$&BTCV dataset. Highly competitive or superior quantitative segmentation results have been achieved using these datasets for four abdominal organs of liver, kidney, spleen and pancreas with reported Dice scores improved by 7.3% for kidneys and 9.7% for pancreas, while being ∼7 times faster than two strong baseline segmentation methods (nnUNet and CoTr).

## 1 Introduction

Automatic abdominal multi-organ segmentation is crucial for the down-streamed clinical tasks of computer-aided diagnosis, imaging biomarker measurement and treatment respond [28,26]. In the abdomen, multi-organ segmentation is still a challenge task due to the nature of large variations of patient data. First, the CT intensity contrast level between abdominal organs is relatively low [12].

Second, the heterogeneous lesions and complex spatial structures of the organs cause the diversity of organ morphology. Finally, different CT scanners, different scanning protocols and contrast agents further diversify the intensity distributions of the organs. In the abdominal region, a variety of classical model-based approaches have been proposed to exploit multi-organ segmentation in past few years. For example, statistical shape models [1,15] use the positional relationship between the organs and the shape of the organs in general statistical space as constraints to regularize/optimize the segmentation. However, when the shape distribution of the target organ has too large non-Gaussian deviations with the mean shape, these methods often fail to segment well. Multi-atlas label fusion related work [23,27] first reconstruct a few numbers of atlases with segmentation annotations, then attempt to fit the target image to the original CT image in each of the atlases via nonrigid registration, and finally propagate and fuse the annotation labels for segmentation. Nevertheless, the nonrigid registration of 3D images is time-consuming and requires similar texture patterns between the target image and original images in the atlases to work.

Recently, fully convolutional neural networks (CNN) based methods and CNN-Transformer based techniques have been proposed and applied to addressing this important task [5,30,22]. For example, two stages segmentation methods are reported [30,22] which usually utilize the localization information from the first stage to assist the second stage of multi-organ segmentation tasks. Isensee et al. [8] propose a general CNN-based segmentation framework obtaining the state-of-the-art performances in many medical image segmentation benchmarks and often serving as a strong segmentation baseline. Due to its ability of long-range contextual modelling, transformer is used to medical image segmentation. Chen et al. [2] present a novel network architecture that uses transformer to further encode the CNN encoder for building a stronger encoder and report competitive multi-organ segmentation results. Xie et al. [31] present a CNN-Transformer based technique that incorporates the deformable self-attention mechanism into transformer for reducing the computational and spatial complexities. Although these methods have shown competitive results, they may fail to segment the data which is rarely seen in the training set or the heavily pathological organs.

In this paper, we propose a novel fully-supervised deep learning image segmentation model that provides robust and accurate segmentation results for multiple abdominal organs (Fig. 1a). It is a modified V-Net model [13] combined with an input decomposition module via a conditional variational auto-encoder (cVAE) [6,9,24] and a hierarchical spatial feature transform module for boosting segmentation accuracy and feature representation. 1) The input decomposition module can represent complex distributions and probabilistically encode possible organ-specific segmentation variants to a low dimensional latent space [10]. 2) The hierarchical spatial feature transform (HSFT) module generates conditional *Affine* transformation parameter pairs for each spatial scale based on the sample from the learned latent space, and it can better preserve organ semantic variations against common normalization layers [16]. 3) Different from [29,16], HSFT module does not need segmentation probability maps or segmentation masks as the external data for generating the Affine transformation parameters. By integrating the predicted organ-specific semantic information as prior and modulating the feature maps for each scale, it generalizes well to the data which is rarely representative as seen from the training set or the organs are heavily affected by lesions. We have evaluated our approach on two publicly available datasets. The quantitative experimental results demonstrate that our method is highly competitive or superior than previous state-of-the-art techniques [8,31,2].

## 2   Methods

Our proposed network architecture is a new V-Net model modified with a cVAE as input decomposition module and a HSFT generator module. Fig. 1a shows the overall structures of the proposed network architecture: the prior encoder and posterior encoder have the same architecture that decomposes the inputs into N-dimensional axis-aligned Gaussian latent space (N is set to 64 in our experiments) and probabilistically learns organ-specific semantic variants [10]. During training, the random sample from the posterior is set to the hierarchical spatial feature transform generator, then to each spatial feature transform (SFT) block by convolution, in order to produce the modulation parameter pairs ($\alpha_i$, $\beta_i$) with spatial dimensions at each scale. Finally, this learned transformation effectively propagates the semantic information throughout the decoder of V-Net.

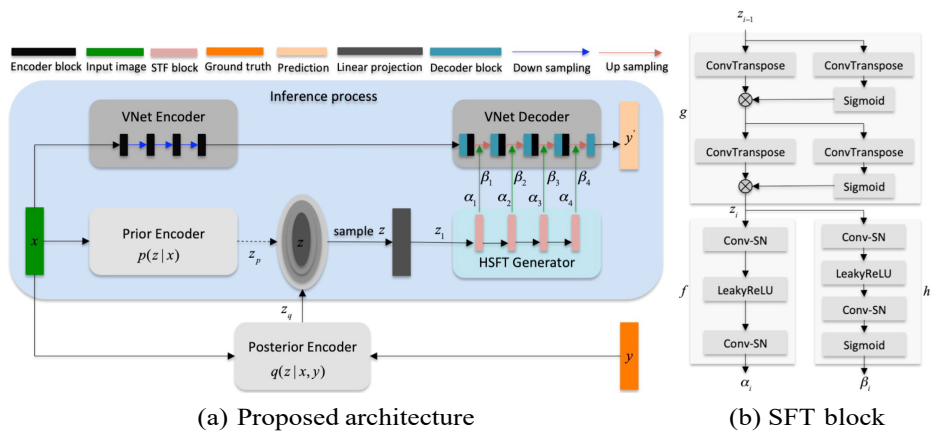

(a) Proposed architecture  (b) SFT block

Fig. 1: The overall network structures of the proposed segmentation method.

We use multi-class cross entropy and generalized dice loss functions [25] as the segmentation loss $L_{Seg}$. We employ Kullback-Leiber divergence to make the prior encoder and conditional posterior encoder approximating each other. When in testing without the ground truth segmentation, the prior encoder still can predict the sample $z_p$ approximately towards the posterior $z_q$ as much as possible, $L_{KL} = D_{KL}(q(z|y,x) \| p(z|x))$. Both losses are combined as a weighted sum with weighting factors $\lambda_1 = 1.0$ and $\lambda_2 = 10.0$ in our experiments,

$$L = \lambda_1 L_{Seg} + \lambda_2 L_{KL} \tag{1}$$

**Input decomposition.** In the abdominal multi-organ segmentation task, the huge diversity or variations of inter-subject and intra-subject is the main challenge. We attempt to construct a low dimensional latent space to encode the possible or all anatomically feasible variants that can be observed in the training dataset to facilitate the downstream tasks. As shown in [10], a cVAE model can capture complex distributions of organ appearance and shape variants. This input decomposition module consists of a prior encoder and a posterior encoder. These two encoders have the same network architecture as composed of four blocks. Each block contains four Conv-ReLU layers, and each layer includes a 3D convolutional layer followed by Rectified Linear Unit (ReLU) activation. Average pooling is used between blocks to increase the receptive field. The organ prior distribution is modeled as axis-aligned Gaussian functions with estimated means and variances, generated by a 1 x 1 x 1 convolution layer. [10] fuses an additional channel obtained from the predicted latent space to the last decoder block of UNet which suits well for single organ segmentation task. However, it cannot extend to multiple organs segmentation because the sample $z$ from latent space without encoding the multi-organ spatial relations. Directly upsampling and tiling the sample $z$ will lead to semantic (infeasible) segmentation errors.

**Hierarchical spatial feature transform (HSFT).** Conditional batch normalization has proven highly effective in many vision tasks [7,4,17,18]. Inspired by previous studies on feature normalization [29,16], HSFT module is designed to apply Affine spatial transformation hierarchically on whose parameters from the predicted organ-specific sample $z$, which enables each V-Net decoder block to incorporate semantic information for guiding segmentation in a coarse-to-fine manner. Different from [29,16], our HSFT module does not need segmentation probability maps or segmentation masks as the external data for generating the Affine transformation parameters (consisting of 4 SFT blocks). More formally, each SFT block learns three functions $f_i$, $h_i$ and $g_i$ based on the predicted prior $z_{i-1}$, where function $g_i$ is used to upsample the prior $z_{i-1}$ as the input of $f_i$, $h_i$ and the next SFT block; and $f_i$ and $h_i$ can be arbitrary functions (such as neural networks). They generate the Affine transformation parameter pairs $\alpha_i$ and $\beta_i$, respectively.

$$z_i = g_i(z_{i-1}), \qquad \alpha_i = f_i(z_i), \qquad \beta_i = h_i(z_i) \tag{2}$$

As shown in Fig. 1b, function $g_i$ consists of two gated transposed convolutional layers. Function $f_i$ contains two Conv-SN layers, and each layer includes a 3D covolutional layer followed by spectral normalization (SN) [14], and Leaky Rectified Linear Unit (LeakyReLU) is used between these two layers. The architecture of function $h_i$ is the same with function $f_i$, except that the sigmoid layer as the output layer is applied.

After obtaining the Affine transformation parameters from the organ-specific priors, the operator of modulation is performed. More specifically, the Affine transformation parameters are multiplied and added to each decoder block in the V-Net architecture, in both feature-wise and spatial-wise manipulations. It restricts the prior information stored in $z$ to only refine the specific organs of the input images.

$$F_i = \alpha_i \otimes F_i + \beta_i \tag{3}$$

where $F_i$ is the output of V-Net $i^{th}$ decoder block, $\otimes$ is the element-wise multiplication.

## 3 Experiments and Results

**Training dataset:** AbdomenCT-1K contains 1112 abdominal CT scans by incorporating and extending several existing benchmark datasets of LiTS[1], KiTS[2], MSD[3] and NIH Pancreas[4]. These CT scans are from 12 medical centers with multi-phase, multi-vendor and multi-disease patient cases [12]. All cases are provided with the annotated liver, left and right kidney, spleen and pancreas masks and the re-annotated existing benchmark datasets termed as LiTS+, KiTS+, MSD+ and NIH+. The fully supervised abdominal organ segmentation benchmark[5] builds two subtasks based on different collections of 361 training cases from AbdomenCT-1K. Training set of subtask 1 is composed of MSD+(281 cases) and NIH+(80 cases) with all portal phase CT scans. Training set of subtask 2 is composed of MSD+(281 cases), KiTS+(40 cases) and LiTS+(40 cases).

**Testing dataset:** Our proposed approach is evaluated on two publicly available testing datasets. The first set contains 100 challenging (mostly pathological) patient cases from the AbdomenCT-1K fully supervised abdominal organ segmentation benchmark. The other testing dataset has 90 abdominal CT scans in total: 43 scans are re-annotated by [5] from the Cancer Imaging Archive Pancrease-CT(termed as TCIA+) [21,20,3] and 47 scans are from Beyond the Cranial Vault(BTCV) [11]. Different from the first testing dataset, the TCIA+& BTCV[6] set evaluates only liver, left kidney, spleen and pancreas.

### 3.1 Implementation Details

For pre-processing, we first crop the CT scans by thresholding to get a rough abdominal 3D region of interest (ROI), then clip the voxel intensity value to the soft tissue CT window range [-175, 275] Hounsfield Units (HU) to remove the irrelevant details and scale to [0, 1] linearly. The target ROI is resized to 128 x 128 x 128 voxels. In the training stage, we employ RandomBlur, RandomGamma and Ran- domNoise from TorchIO Library [19] to do online intensity augmentation

---

with
[1] https://competitions.codalab.org/competitions/15595
[2] https://kits19.grand-challenge.org/
[3] http://medicaldecathlon.com/
[4] https://wiki.cancerimagingarchive.net/display/Public/Pancreas-CT
[5] https://abdomenct-1k-fully-supervised-learning.grand-challenge.org/
[6] http://doi.org/10.5281/zenodo.1169361

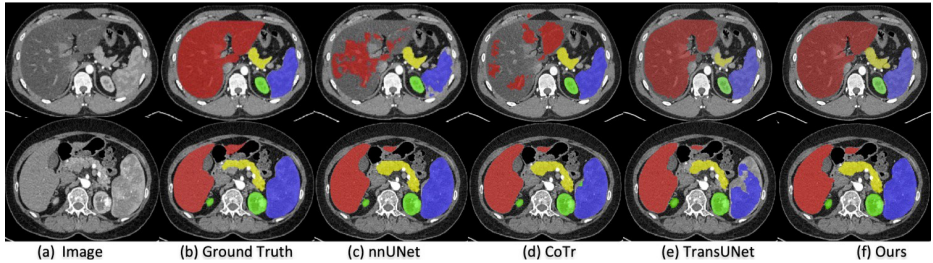

Fig. 2: Segmentation examples of different methods on AbdomenCT-1K fully supervised abdominal organ segmentation benchmark for each subtask. First row shows the segmentation results of different methods on subtask 1 training set. Second row presents the segmentation results of different methods on subtask 2 training set. The ground truth is from [12].

default parameters. According to the preset probability of each transform, we apply only one of the given transforms at a time. We adopt Adam optimizer with the weight decay of 1e 8 and an initial learning rate of 1e 4 decayed by 10% every 20 epochs. The training batch size is set as 4 and we trained 200 epochs in total. A NVIDIA V100 GPU with 32 GB memory is used for training. For inference, we process the testing cases following the above pre-processing steps. Then we resize the segmentation prediction from 128 x 128 x 128 to the size of the cropped abdominal ROI and fill it back to the original volume size.

### 3.2  Results

Our proposed abdomen multi-organ segmentation method is evaluated on Abdomen CT-1K fully supervised abdominal organ segmentation benchmark (Liver, left and right Kidneys, Spleen and Pancreas) and TCIA[+]&BTCV dataset (Liver, left Kidney, Spleen and Pancreas) of four organs in each subtask respectively. We employ the Dice coefficient scores (in percentages %) as the quantitative evaluation metric with standard deviations reported. To further evaluate the performance of our proposed approach, we compare it to the commonly representative CNN based architecture of [8] which leading many benchmarks on multi-organ segmentation, and compare to two other CNN-Transformer architectures of [2] and [31].

**Performance on AbdomenCT-1K:** Table 1 gives the quantitative segmentation evaluation results on each organ for each subtask. It can be observed that our method overall achieves very competitive segmentation accuracy performance against three strong baselines [8,2,31] in subtask 1. In particular, our model produces average Dice gains of 7.3% and 9.7% than [8] for kidneys and pancreas respectively from subtask 1. In subtask 2, the performance of four methods on liver, kidney and kidneys are almost the same or very comparable. Our method's performance on pancreas segmentation is 3.5% lower than [8] in Dice coefficient, but with the lowest standard deviation 14.9% reported. Fig. 2 shows qualitative segmentation results of two challenging patient cases (listed

Table 1: Quantitative comparison for each subtask on AbdomenCT-1K fully supervised abdominal organ segmentation benchmark.

| Training | Methods | Liver (%) | Kidneys (%) | Spleen (%) | Pancreas (%) |
|---|---|---|---|---|---|
| MSD[+](281) NIH[+](80) Subtask 1: 361 | nnUNet [8] | **95.8±6.04** | 84.1±14.8 | 89.9±15.5 | 65.0±22.7 |
| | TransUNet [2]* | 95.3±3.46 | 83.9±17.0 | 91.7±11.9 | 62.2±23.1 |
| | CoTr [31]* | 95.6±5.65 | 83.5±16.8 | 90.6±12.1 | 69.5±21.2 |
| | Ours | 95.3±4.74 | **91.4±11.4** | **92.0±15.2** | **74.7±14.2** |
| MSD[+](281) LiTS[+](40) KiTS[+](40) Subtask 2: 361 | nnUNet [8] | **97.0±2.93** | **91.7±11.6** | 93.6±13.3 | **78.1±15.8** |
| | TransUNet [2]* | 95.5±3.35 | 87.8±13.8 | 91.8±14.6 | 64.4±22.5 |
| | CoTr [31]* | 96.6±3.55 | 90.8±11.3 | **93.7±13.2** | 76.4±17.4 |
| | Ours | 96.2±2.06 | 91.2±12.6 | 92.6±13.9 | 74.6±14.9 |

*The results of two methods were not given on this benchmark, we conduct the experiments (trained 1000 epochs) based on the open source code which is provided by the related papers [2,31].

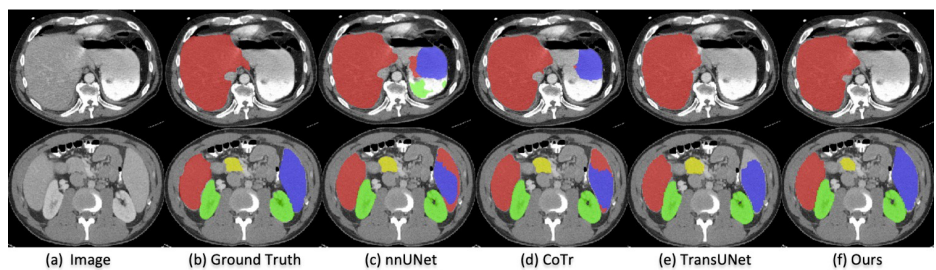

Fig. 3: Segmentation examples of different methods on TCIA[+]&BTCV for each subtask. First row shows the segmentation results of different methods on subtask 1 training set. Second row presents the segmentation results of different methods on subtask 2 training set.

in [12] from each subtask). The first row in Fig. 2 presents a case with fatty liver where [8] and [31] cause severe under-segmentation; our model and [2] generate evidently more complete liver in subtask 1. The second row shows another case's segmentation result in subtask 2. Our model can segment spleen completely, whereas spleen is under-segmented using [8,2] and [31] exists incorrect segmentation (kidney mask on Spleen).

**Performance on TCIA[+]&BTCV:** Table 2 provides the numerical segmentation results of our models and three comparison methods of each organ in each subtask. Except for pancreas, our approach achieves quantitatively the best accuracy (averaged Dice) and most stable (standard deviation) segmentation results than other methods [8,2,31]. Fig. 3 illustrates the segmentation results from each subtask on TCIA[+]&BTCV. In subtask 1 (first row), it can be seen that our model generates the correct and more complete segmentation result than all other three methods. The second row shows a case's segmentation result in subtask 2, only our model is able to segment right kidney completely.

**Discussion** In contrast from [8] and [31], our approach is designed based on the learned semantic variants via a probabilistic cVAE model and learned organ-specific semantic variations being integrated into each V-Net decoder block by hierarchical spatial feature transformation for segmentation regularization. So even if some testing sets have severe diseases and demonstrate low image quality (where the training sets have very few similar cases), our model achieves quantitatively competitive segmentation results for each organ in each evaluation subtask. Especially in subtask 1, the training set is all portal phase CT, which is a pancreas related subset from AbdomenCT-1K, our method can do well by adequately generalizing to the lesion-affected liver, kidney and spleen organs, it also handles well multi-phase CT imaging data in testing sets (shown as the first row in Fig. 2). From Table 1 and Table 2, we observe that our approach obtains more statistically stable segmentation performance with different training sets. However, [8] and [31] have shown better performance in pancreas segmentation (in Table 2). In order to achieve very high inference efficiency, we downsample the input to 128 128 128, which likely affecting the spatially narrow type of organ segmentation such as pancreas. In unseen testing cases where lesions could have severely affected the shape and appearance of the organs, both our approach and several state-of-the-art methods [8,31,2] have challenges on handling multi-organ segmentation accurately and reliably (Appendix A).

Table 2: Quantitative comparison for each subtask on TCIA[+]&BTCV.

| Training | Methods | Liver (%) | Left Kidney (%) | Spleen (%) | Pancreas (%) |
|---|---|---|---|---|---|
| MSD[+](281) NIH[+](80) Subtask 1: 361 | nnUNet [8]* | 95.4±4.17 | 89.9±10.7 | 93.5±7.62 | **83.0±4.71** |
| | TransUNet [2]* | 93.85±11.2 | 87.8±15.3 | 91.4±10.9 | 70.8±17.8 |
| | CoTr [31]* | 95.3±4.57 | 89.8±11.1 | 92.8±9.02 | 82.2±5.42 |
| | Ours | **96.1±0.84** | **90.2±10.2** | **95.4±1.45** | 75.8±14.7 |
| MSD[+](281) LiTS[+](40) KiTS[+](40) Subtask 2: 361 | nnUNet [8]* | 95.7±3.86 | 91.7±10.2 | 94.5±6.96 | **82.6±5.07** |
| | TransUNet [2]* | 94.4±10.1 | 89.8±13.7 | 93.5±7.96 | 73.5±10.6 |
| | CoTr [31]* | 95.7±4.37 | 91.7±10.1 | 94.5±7.67 | 82.4±5.11 |
| | Ours | **95.9±0.94** | **91.8±9.98** | **95.4±3.51** | 78.4±9.09 |

*The comparison results of Subtask 1 are from only BTCV dataset and Subtask 2 are from TCIA[+]&BTCV.

## 4 Conclusion

In this paper, we present a new probabilistic V-Net model with hierarchical spatial feature transform for abdominal multi-organ segmentation. Our key idea is that we use a cVAE module conditioned on input images to learn the organ-specific distributions on low dimensional latent space and capture richer semantic organ appearance variations. The learned semantic information is injected into each V-Net decoder block hierarchically by each spatial feature transform block for boosting the multi-organ segmentation performance in both accuracy and inference efficiency. The proposed approach has been quantitatively evaluated on two public datasets, demonstrating robust generality and high accuracy to handle lesion-affected/pathological organs and cope with low image quality cases.

## Appendix Challenging segmentation cases

**Challenging Segmentation Cases with Failures**

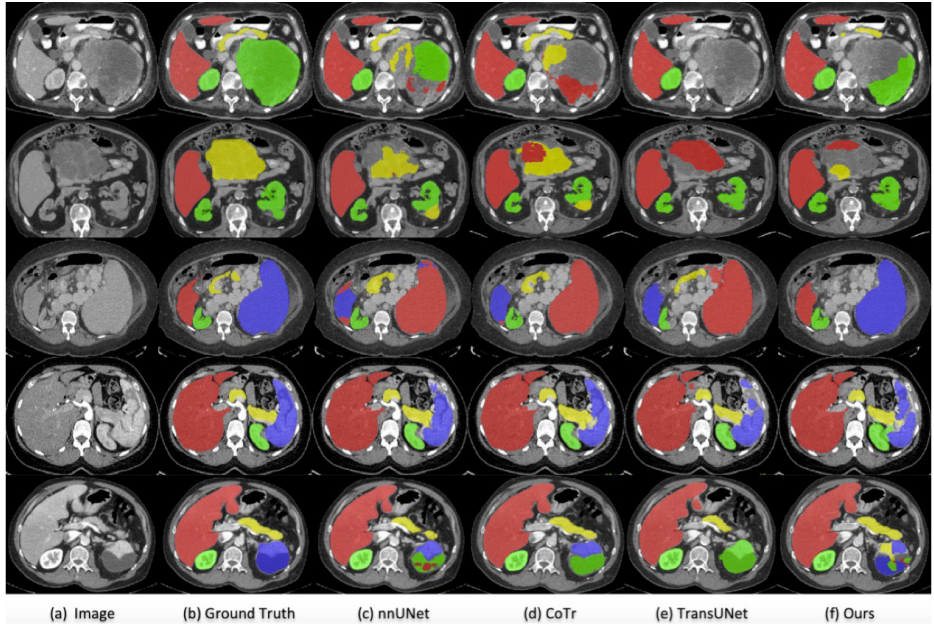

Fig. 4: Challenging segmentation examples with failures for all four different methods per subtask. The first row presents the segmentation results of four methods on the subtask 1. The second row show the segmentation results on the subtask 2. Case 3 is from TCIA$^+$&BTCV testing set, other two cases are from the testing set of AbdomenCT-1K fully supervised abdominal organ segmentation benchmark. Abdominal Multi-Organ segmentation task in CT imaging for long-tailed (often pathological) patients is still an unsolved problem, pending further research efforts.

**Challenging but Successful Segmentation Cases**

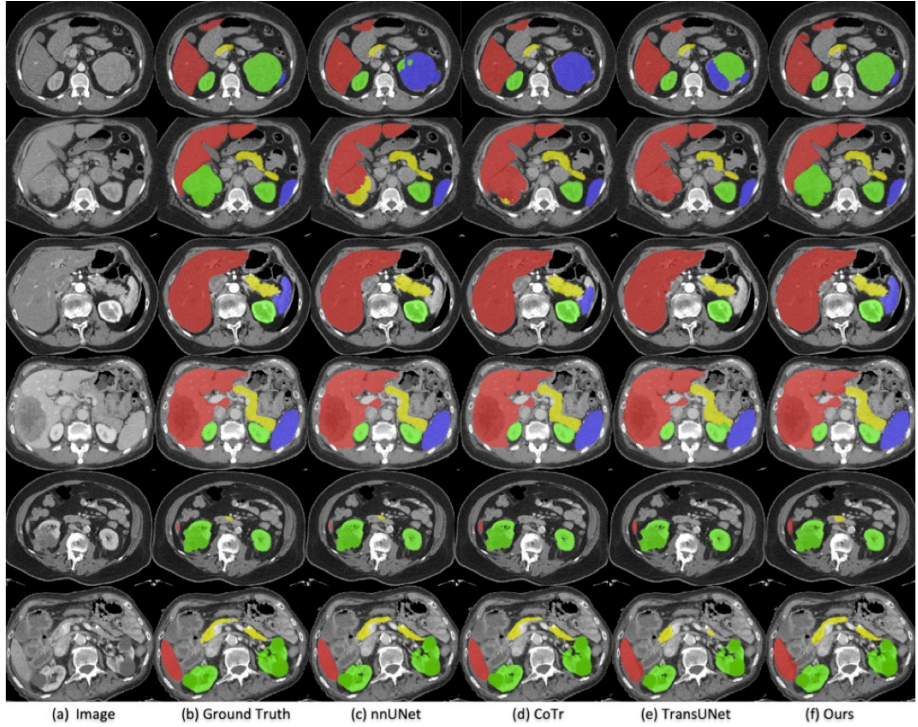

Fig. 5: Challenging yet successful segmentation examples from the testing sets of AbdomenCT-1K fully supervised abdominal organ segmentation benchmark for each subtask. The top three rows present the segmentation results of four different methods on the subtask 1. The bottom three rows illustrate the segmentation results of four methods on the subtask 2.

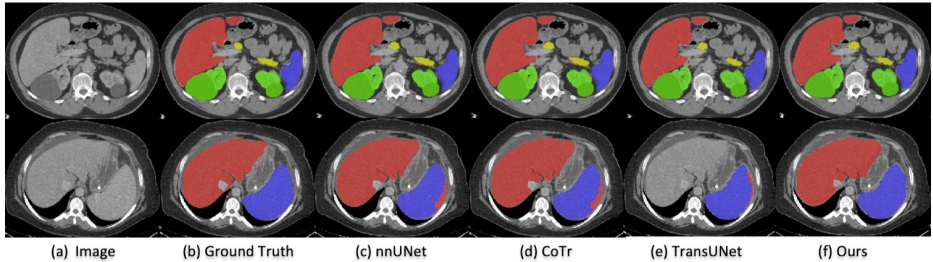

Fig. 6: Successful segmentation examples of four different methods on TCIA[+]&BTCV dataset for each subtask. The first row presents the segmentation results of four methods on the subtask 1. The second row shows the segmentation results of four methods on the subtask 2.